\documentclass[letterpaper,aps,pra,nolongbibliography,twocolumn,showpacs,floatfix,10pt]{revtex4-1} 
\usepackage{graphicx,graphics}	
\graphicspath{{images/},{images_Tesis/},{"/home/david/Escritorio/Maestria/spin chain/Moviendobz/"}}
\usepackage[utf8]{inputenc}
\usepackage{eqnarray}
\usepackage{tikz}
\usepackage{bm}

\usepackage[draft,inline,nomargin]{fixme} \fxsetup{theme=color}

\expandafter\let\csname equation*\endcsname\relax 
\expandafter\let\csname endequation*\endcsname\relax
\usepackage{amsmath,amssymb,bbm} 
\usepackage{color}
\def\notext#1{}

\newcommand{\mcF}{\mathcal{F}}	
\newcommand{\mcH}{\mathcal{H}}	
\newcommand{\env}{\text{env}}	
\renewcommand{\env}{{\rm env}}	
	
\newcommand{\sys}{\text{sys}}

\newcommand{\mcJ}{\mathcal{J}}

\newcommand{\mcE}{\mathcal{E}}
\newcommand{\mcG}{\mathcal{G}}
\newcommand{\rhp}{RHP}	
\newcommand{\mcD}{\mathcal{D}}

\newcommand{\rmf}{f}

\newcommand{\<}{\langle}
\renewcommand{\>}{\rangle}
\def\1{{\mathchoice{\rm 1\mskip-4mu l}{\rm 1\mskip-4mu l}{\rm 1\mskip-4.5mu l}{\rm 1\mskip-5mu l} }}

\newcommand{\nm}{NM}
\newcommand{\ipr}{IPR}

\newcommand{\mcN}{\mathcal{N}}

\newcommand{\mcK}{\mathcal{K}}

\newcommand{\ising}{V_J}
\newcommand{\localising}{V_{0,1}}
\newcommand{\field}{V_b}
\newcommand{\localfield}{V_0}
\newcommand{\blp}{BLP}
\newcommand{\rmd}{\text{d}}
\newcommand{\fref}[1]{Fig.~\ref{#1}} 
\newcommand{\eref}[1]{Eq.~(\ref{#1})} 
\newcommand{\sref}[1]{Sec.~\ref{#1}}

\newcommand{\ie}{i.e.}

\newcommand{\ket}[1]{{\vert #1 \rangle}}

\newcommand{\proj}[2]{{\vert #1 \rangle \langle #2 \vert}}

\newif\ifenglish
\newif\ifespanol
\newif\ifshort
\newif\iflong
\shorttrue
\englishtrue

\usepackage{hyperref}

\providecommand{\openone}{\leavevmode\hbox{\small1\kern-3.8pt\normalsize1}}
\begin{document}
\title{Quantum non-Markovianity and localization} 
\author{David Davalos$^1$, Carlos Pineda$^{1,2}$}
\affiliation{$^1$Instituto de Física, Universidad Nacional Autónoma de México, México D.F. 01000, México}
\affiliation{$^2$University of Vienna, Faculty of Physics, Boltzmanngasse 5, 1090 Wien, Austria}
\date{\today}%
\begin{abstract}
We study the behavior of non-Markovianity with respect to the localization of
the initial environmental state. The ``amount'' of non-Markovianity is measured
using divisibility and distinguishability as indicators, employing several
schemes to
construct the measures.
The system used is a qubit coupled to an environment modeled by an Ising spin
chain kicked by ultra-short pulses of a magnetic field.
In the integrable regime, non-Markovianity and  
localization do not have a simple relation, but as the chaotic regime is
approached, simple relations emerge, which we explore in detail. 
We also study the non-Markovianity measures in
the space of the parameters of the spin coherent states and point out that
the pattern that appears is robust under the choice of the interaction
Hamiltonian but does not have a classical-like phase-space structure.
\end{abstract}
\pacs{03.65.Yz,03.67.-a}
\maketitle
\section{Introduction}  
Open quantum systems were recognized as an important subfield of quantum
mechanics early in their history~\cite{Neumann1927}, because
understanding them allows one to explain ubiquitous phenomena, such as
spontaneous decay~\cite{Dirac243}.  Later, the Lindblad equation  was proposed to describe the evolution of the reduced density
matrix of a quantum system weakly coupled to a memoryless
environment~\cite{lindblad,kossa,gorini}. 
Environments that lie outside that approximation 
(Lindblad equation) have attracted
the attention of the community in later years. This is, arguably, because we
now have such delicate control of quantum systems that memory effects become
experimentally relevant~\cite{RevModPhys.89.015001}, and environment
engineering is possible~\cite{Verstraete2009,Nokkala2016} to mitigate or even
use such effects~\cite{rivasreview,RevModPhys.88.021002,RevModPhys.89.015001}.
A whole community is now dedicated to the study of such systems, known as {\it
non-Markovian} environments. Numerous efforts 
have been made to define non-Markovianity (\nm{}) in a precise manner, to measure it, and to
take advantage of it (see the previous review papers
and~Refs.\cite{ourmeasure, Poggi2017}). Many systems have been studied under this
program, both theoretically and experimentally~\cite{RevModPhys.89.015001}.\par

Currently, there are many examples of non-Markovian environments that produce a
variety of effects. However, not much is known regarding what the key
properties that might boost the non-Markovianity of an environment are.  
Some properties, such as the structure of the phase space of the classical
counterpart of the environment have proven to be crucial; however, 
what happens when we do not find such a classical analog?
In this paper we focus on two questions. First, is the value of the several
measures of non-Markovianity for long times, only dependent on the effective
dimension of the Hilbert space?  Second, is there a hidden underlying classical
structure in the environment that we can unveil with the help of these
measures?\par

To study these questions, we consider a qubit coupled 
to a kicked spin chain, which has integrable, mixed and chaotic dynamical
regimes~\cite{prosen,prosen2000}, but, as far as we know, no semiclassical
analog. 
The interaction between qubit and environment is set up so as to have
dephasing, so all the decoherence effects on the qubit are contained in
a suitably defined fidelity of the environment.  
To quantify \nm{}, we use two commonly used measures
\cite{breuermeasure,rhp} and a third that was recently introduced and which
has a direct relation with a physical task~\cite{ourmeasure}. 

We find complex relations between \nm{} and the localization of
initial environmental states in the integrable and mixed regimes, which depend
on the peculiarities of each \nm{} measure. 
In fact, in Ref.\cite{Lorenzo2017} a relation between localization, induced
by disordered, and a particular non-Markovianity measure was explored for an
environment consisting of an array of cavities. 
In the case of the recently
introduced measures~\cite{ourmeasure}, the effective dimension of the Hilbert
space of the environmental states has an important role which leads to 
more complex behavior. In the chaotic regime, due to the ergodic properties
of the Hamiltonian, the relation is simpler and almost homogeneous.
Regarding the search for underlying classical structure, we focus our attention
on the features that emerge in the space of
the parameters of the initial states (spin coherent states) when the \nm{ } and
the \textit{inverse participation ratio} (\ipr{}) \cite{seligmanipr} are
calculated. 
%
We searched for the characteristic finely granulated 
fractal structure predicted   by the
Kolmogorov–Arnold–Moser (KAM) theorem 
but found only a coarse non fractal one.

\par
The paper is organized as follows. In Sec.~\ref{sec:tools} we give a brief
introduction to the measures used for non-Markovianity and for localization of
quantum states. In Sec.~\ref{sec:model} we present the general scheme of
dephasing dynamics and the details of the dynamics.
In Sec.~\ref{sec:results}, we present and discuss the results. We finish 
by summarizing the results in Sec.~\ref{sec:conclusions}.
\section{Tools} 
\label{sec:tools}
\subsection{Identifying non-Markovianity} 
Many measures of non-Markovianity have been proposed: The two most wide spread
are the BLP (introduced by Breuer, Laine and Piilo in~\cite{breuermeasure}) and
RHP (introduced by Rivas, Huelga and Plenio in~\cite{rhp}) measures. The first
is
based on the violation of the contraction property of Markovian systems, \ie{},
decreasing  distinguishability between initial quantum states. The second is
based on the violation of a well known mathematical property of Markovian
process, divisibility of the quantum map. Both criteria come from the classical
theory of Markovian stochastic process. A whole new set of measures have 
been proposed~\cite{RevModPhys.88.021002}. One of these~\cite{ourmeasure}, proposed by
the authors of this paper, is based on quantifying the probability of 
successfully performing a certain task. 
%
%

It is hard to strictly verify if a stochastic system fulfills the classical 
definition of Markovianity \cite{markovlife}, since it depends on the whole
history of the stochastic process. An additional caveat for quantum systems is
the fact that in order to observe intermediate states of the system, one would
have to measure, thus collapsing the wave function and thus also the
probability distributions.  This leads,  among other problems, to violation of
Kolmogorov consistency conditions even for closed quantum systems
\cite{RevModPhys.88.021002}.  
\par
\begin{figure} 
\centering
\includegraphics{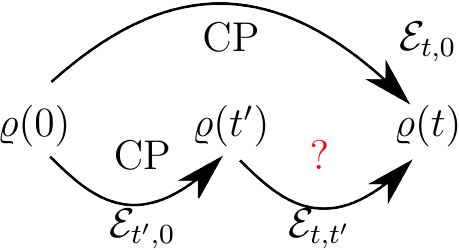}
\caption{Illustration of the concept of CP divisibility. 
The process $\mcE$ is CP divisible if
all existing intermediate maps $\mcE_{(t,t')}$,
are complete positive and trace preserving.
\label{flechas}}
\end{figure}

%
One can, however, check the necessary conditions for Markovianity that can be
easily interpreted from a physical point of view. For example notice that a
classical stochastic process (not necessarily Markovian) can be described by a
time dependent right stochastic matrix $A(t)$ that maps the initial probability
distribution $\vec p(t=0)$ to  $A(t)\vec p(0)=\vec p(t)$.  Matrices describing
the intermediate process, say the map from time $t'$ to $t\ge t'\ge0$, described by 
$A_{t,t'} \equiv A_{t,0}A_{t',0}^{-1}$, will also be right stochastic matrices
for Markovian processes. We argue that the intermediate process is a
valid one, and if $A_{t,t'}$ is right stochastic for all $t\ge t'\ge0$,
the process is said to be divisible.
This construction can extended to the
quantum case, replacing the divisibility concept with the {\it completely 
positive} map (CP map), which characterizes a valid quantum channel. Given 
a quantum process $\mcE_{t,0}$, we shall say that it is CP divisible
if the intermediate dynamics 
\begin{equation}
\mcE_{t,t'}\equiv 
   \mcE_{t,0}\mcE^{-1}_{t',0}, \hspace{10 pt} t \geq t' \geq 0
\label{eq:divisibility}
\end{equation}
are CP maps. 
Figure~\ref{flechas} illustrates the general idea for divisibility and
CP divisibility.
A general property of a CP divisible process is that given any Hermitian operator
$\Delta$ the trace norm decreases under the action of the
map~\cite{rivasreview}
$\left| \left| \mcE \left(\Delta\right) \right| \right|_1\leq \left| \left| \Delta \right|\right|_1$, where 
$\vert \vert \cdot \vert \vert_1$ is the \textit{trace norm}. In particular,
choosing $\Delta=1/2\left( \varrho_1-\varrho_2 \right)$ we have
\begin{equation}
D\left(\mcE\left(\varrho_1\right),\mcE\left(\varrho_2\right)\right) \leq
D\left( \varrho_1,\varrho_2 \right)
\label{eq:distance}
\end{equation}
where $D(\varrho_1,\varrho_2)=1/2 \vert\vert \varrho_1-\varrho_2 \vert\vert_1$
is the \textit{trace distance}.
This property shows the contraction of the state space under a Markovian
process.
This in turn shows how two initial conditions are increasingly forgotten, 
and are more difficult to distinguish, as the trace norm is 
directly related with the two state discrimination problem. 
Some authors {\it define} Markovianity with this property: 
If there exists a pair of quantum states such that the last equation does not
hold, in Ref.~\cite{breuermeasure} the process is said to be non-Markovian.
\par
\subsection{Quantifying non-Markovianity} 
\label{sec:measures}
Two well-known measures of non-Markovianity can be constructed, based on
violations of either Eqs.~(\ref{eq:divisibility}) or (\ref{eq:distance}). In
particular, the authors of both measures constructed them adding up the local
contributions of the chosen criterion. \par
For the case of the RHP measure (based on divisibility) the authors define 
\begin{equation}
g(t) = \lim_{\epsilon \to 0^+} \frac{\left| \left| \mcJ[\mcE_{(t+\epsilon,t)}] \right|\right|_1 -1}{\epsilon},
\label{eq:goft}
\end{equation}
where $\mcJ[\mcE_{(t+\epsilon,t)}]$ is the Jamiołkowski
isomorphism~\cite{bengtsson2006geometry} that relates quantum
channels and density matrices. In particular, it takes
CP maps to positive operators with unit trace. Thus, if 
$\mcE_{(t+\epsilon,t)}$ is a CP map, the eigenvalues of the 
$\mcJ[\mcE_{(t+\epsilon,t)}]$ will all be positive and 
add up to one. Otherwise, they will still add up to one, but with
negative contributions. Thus, 
$g(t)$ is greater than zero if  at time $t$ the dynamics are not
divisible; otherwise, $g(t)=0$. 
The measure proposed in Ref.~\cite{rhp} is obtained 
by integrating the contributions of the non-CP-divisible behavior throughout the entire evolution:
\begin{equation}
\mcN_{\text{RHP}}\left[\mcE\right]=\int_0^{\infty} g(t) \rmd t.
\label{eq:rhp}
\end{equation}
The brackets here indicate functional dependency.\par
In a similar spirit, we can integrate the deviations from 
the contractive behavior, expected for Markovian evolution. 
Considering the derivative of the trace distance
\begin{equation}
\sigma \left( t, \varrho_{1,2}(0)\right) = 
\frac{\rmd D \left( \varrho_1(t),\varrho_2(t) \right)}{\rmd t}.
\end{equation}
According to \eref{eq:distance}, $\sigma \leq 0$ for Markovian 
dynamics. We can integrate this deviation to obtain the 
measure proposed in Ref.~\cite{breuermeasure}, where a maximization 
over all states is taken. Thus, 
\begin{align}
\mcN_{\text{BLP}}[\mcE] &= \max_{\varrho_1,\varrho_2} \int_{\sigma>0} \sigma \left( t, \varrho_1(0),\varrho_2(0) \right) \rmd t . 
\label{eq:blp}
\end{align}
\par 

These two measures have some serious drawbacks. In particular, they are not
continuous in the spaces of functions, and small fluctuations can 
change the value of the measure by an arbitrarily large amount. 
Notice that these issues arise always with a finite Hilbert size environment, 
and also in finite number statistics.
One has
the option to cut the integration interval to a finite time, or 
smooth out the fluctuations by windowing the data. One can also 
consider other proposals~\cite{ourmeasure}  which not only remove
that problem, but also provide a physical interpretation for 
the number obtained. The proposals are 
\begin{equation}
\mcN^{\max}_\mcK[\Lambda_t ] = \max_{t_f,\tau \le t_f }
\left[ K(t_\rmf) - K(\tau) \right]
\label{eq:GorinNM}
\end{equation}
and
\begin{equation}
\mcN^{\<\cdot \>}_\mcK[\Lambda_t ] =
\max \left\{ 0,
\max_{t_f }
\left[
K(t_f) - \< K(\tau) \>_{\tau<t_f}
\right]
\right\}.
\label{eq:CarlosNM}
\end{equation}
In this case, $K$ is a quantity associated with the channel and/or its
derivative. This can be,  say, the quantum
capacity, the trace distance with respect to some fixed states, 
or even $\dot K(t)=g(t)$ as defined in \eref{eq:goft}.
\subsection{Fidelity and localization} 

A very simple model of an open quantum system is one in which the 
dynamics of both the system of interest (central system) and 
environment are considered and taken to be unitary. If the
interaction between them commutes with the Hamiltonian 
governing the system, one has dephasing dynamics. This kind
of dynamics is the simplest decoherence type and is the 
one considered in this article. If the central system is a 
qubit, one can write the evolution operator as
\begin{equation}
U=|0\>\<0|\otimes U_0 + |1\>\<1| \otimes U_\delta
\label{eq:unitary:dephasing}
\end{equation}
with $U_0$ and $U_\delta$ acting on the environment and $|i\>\<i|$ ($i=0,1$)
appropriate
projectors on the qubit.
Given that the initial state of the whole system is the separable state 
$|\psi_\sys \> \otimes |\psi_\env\>$, the dynamics on the qubit only depend on the
{\it fidelity amplitude}~\cite{echo} defined as 
\begin{equation}
f(t)= \<\psi_\env|U_\delta^\dagger(t) U_0(t) |\psi_\env\>
\label{eq:fidelityaplitude}
\end{equation}
and the expectation value of the echo operator $M(t)=U_\delta^\dagger(t) U_0(t)$
with respect to the state $|\psi_\env\>$. In particular, 
the unitary dynamics of the qubit are going to be encoded in the phase 
of $f$; other quantities such as purity, that are invariant under unitary
transformations, depend only on the fidelity 
\begin{equation}
\mcF(t)=|f(t)|^2.
\label{eq:definition:fidelity}
\end{equation}
It follows that in the dephasing scenario, the study of non-Markovianity
reduces to the study of the fidelity amplitude in the environment.

If we consider long discrete times, and under ergodic conditions, one can
assume that the sequence of states $M(t)|\psi_\env\>$ is random with respect to
$|\psi_\env\>$; by that we mean that $\<\psi_\env|M(t)|\psi_\env\>$ is a sequence of
random Gaussian numbers.  In this model the fidelities
are uncorrelated Gaussian random numbers with zero mean and standard deviation
inversely proportional to the square root of the dimension  of the
Hilbert space in which $|\psi_\env\>$ lives.  
However, systems that are not ergodic, from a classical point of view, do not
explore the whole phase space. The simplest correction to the model proposed
leads
to the concept of {\it effective Hilbert space}. The dynamics, for a fixed
initial state, can often be described with smaller subset of states sharing 
a quantum number with the initial state. 
Say, if the initial state of a semiclassical integrable system lives 
in a torus, we can describe the evolution with the eigenstates belonging
to that same torus.
Thus, the dynamics are taking place in 
an effective Hilbert space of dimension  roughly equal to the number 
of coherent states that cover that torus. In a purely quantum scenario, 
such a situation arises naturally when one has ``good'' quantum numbers. A
reasonable way to quantify to what extent one can describe states in
terms of a small number of states of an orthonormal basis is using the
\textit{inverse participation ratio} (\ipr{}). This quantity is defined 
for a normalized state $\ket{\psi}$ with respect to the orthonormal 
basis $\{|n\>\}$ as 
\begin{equation}
\text{P}^{-1}(\ket{\psi})
  =\sum_n^{\dim \mcH} \left| \langle n \vert \psi \rangle \right|^4.
\label{eq:ipr}
\end{equation}
The lower bound for the \ipr{} is $1/\dim \mcH$ and is attained when we have
equal weights of $|n\>$ on the state $\ket{\psi}$; we say that $\ket{\psi}$ is
a fully delocalized state. The upper bound of $1$ is obtained by states of the
base  $\{|n\>\}$; we say that $\ket{\psi}$ is localized.  Typically the basis
$\lbrace \ket{n} \rbrace$ is chosen as the normal eigenbasis of some operator,
typically the Hamiltonian prior to a perturbation. It should be noted that
such an operator can not have degenerate spectra in order to avoid ambiguities in
the basis and get well-defined \ipr{}s.
\subsection{Putting together the tools} 
At this point, we wish to connect the three quantities discussed:
non-Markovianity measures, fidelity, and \ipr{}. Non-Markovianity measures are
determined, for dephasing channels, by the fidelity of an environment. In
particular, as can be seen from Eqs.~(\ref{eq:rhp}) and (\ref{eq:blp}), they are
determined by the fluctuations of fidelity. In turn,  under an ergodic
hypothesis, the \ipr{} can tell us how asymptotic fidelity behaves,
with an effective dimension yet to be determined. In this paper we want
to study under which circumstances we  can reduce the study of non-Markovianity
to the study of an effective dimension of a quantum system.\par
\section{Model} 
\label{sec:model}
In this section we start with a generic Hamiltonian that induces dephasing dynamics. 
We then specify the particular model to be used as environment, namely, a
kicked chain of spin-$1/2$ particles and the initial states of the environment. 
We complete our model specifying the interactions considered in this work. 

\subsection{Dephasing dynamics} 
The Hamiltonian of a qubit under dephasing dynamics is, up to rotations in the qubit,
\begin{equation}
H  =  \frac{\Delta}{2} \sigma_z \otimes \mathbbm{1}
      + \mathbbm{1} \otimes H_{\text{env}} 
      + \epsilon \sigma_z \otimes V 
\label{eq:ham}
\end{equation}
[as in \eref{eq:unitary:dephasing}, when writing tensor products, the first term 
acts on the qubit and the second, on the environment].
The first term is the free Hamiltonian of the qubit and $\Delta$ is the
transition energy between the two levels; $H_{\text{env}}$ is the environmental
Hamiltonian; finally, $\epsilon$ modulates the coupling strength of the
qubit-environment system, provided by the last term. 
Since the internal Hamiltonian of the qubit commutes with the interaction Hamiltonian we can ignore the latter; it contributes with a unitary transformation in the 
qubit that does not affect the non-Markovianity measures. 
The total Hamiltonian can thus be written as
\begin{equation}
H=\proj{0}{0}\otimes H^{(+)}+\proj{1}{1}\otimes H^{(-)},
\label{eq:hamsim}
\end{equation}
where $H^{(\pm)}=H_{\text{env}}\pm \epsilon V$; its associate unitary operator
takes the form \eref{eq:unitary:dephasing}.
If we write the channel in the Pauli basis $1/\sqrt{2} \lbrace \mathbbm{1},
\sigma_x, \sigma_y, \sigma_z \rbrace$, its matrix elements are given by 
$\mcE_{jk}=(1/2)\text{tr}\left[ \sigma_j U(t) \sigma_k
\otimes \varrho_{\text{env}} U^{\dagger}(t) \right]$, where
$|\psi_\env\>\<\psi_\env|$ is the initial state of the environment and
$\sigma_0 \equiv \mathbbm{1}$. We arrive to the expression
\begin{equation}
\mcE=\left(\begin{matrix}
1 & 0 & 0 & 0 \\ 
0 & \operatorname{Re}[f(t)] & \operatorname{Im}[f(t)] & 0 \\ 
0 & \operatorname{Im}[f(t)] & \operatorname{Re}[f(t)] & 0 \\ 
0 & 0 & 0 & 1
\end{matrix} \right)
\label{eq:channel}
\end{equation}
%
%
with $f$ the fidelity of $|\psi_\env\>$ with respect to the unitary operators
$U^{+}(t) = \exp \left( -i t H^{+}\right)$ and $U^{-}(t) = \exp \left( -i t H^{-}\right)$.

%

For this channel, all measures of non-Markovianity given in the last section can be
easily computed and depend only on $F(t)=\sqrt{\mcF(t)}$. For example, 
\begin{equation}
\mcN_{\text{RHP}}\left[\mcE\right]=\int_{\dot F>0} \frac{\dot F(t)}{F(t)} \rmd t
= \sum_{i} \left[ \log\left(F(b_i)\right)- \log\left(F(a_i)\right)  \right],
\label{eq:rhppractical}
\end{equation}
with $b_i$ and $a_i$ the times of the $i$-th maximum and minimum of $F(t)$
respectively. For the computation of the BLP measure, the states that maximize
\eref{eq:blp}
are those lying on the equator of the Bloch sphere in antipodal positions.
The trace distance is the Loschmidt echo, $D(\varrho_1(t),\varrho_2(t))=F(t) $. 
From \eref{eq:blp}, the measure is
\begin{equation}
\mcN_{\text{BLP}}\left[ \mcE \right] 
= \int_{\dot F>0} \frac{\rmd F(t) }{\rmd t}\rmd t
= \sum_{i} \left[ F(b_i)- F(a_i)  \right],
\end{equation} 
This shows a direct relation with both revivals and fluctuations of the
Loschmidt echo of the environmental dynamics. Finally, measures 
$\mcN^{\max}_\mcK[\Lambda_t ] $ and
$\mcN^{\<\cdot \>}_\mcK[\Lambda_t ]$, as long as they are invariant 
with respect to unitary operations in the qubit, will depend only 
on $F$ in the same way that the particular $\mcK$ chosen depends on 
$F$.

%
\subsection{The environment} 
The system used as environment is the homogeneous Ising
spin-$1/2$ chain kicked by short pulses of magnetic field. This system 
was proposed by Prosen to study the relation between ergodicity and
fidelity~\cite{prosen,prosen2000}. The
Hamiltonian reads
\begin{equation}
H_{\text{env}}=\sum_{i=0}^{N-1} \sigma^z_i \sigma_{i+1}^z 
  +\hat  \delta(t) \sum_{i=0}^{N-1} b^{ \perp }\sigma_i^x +b^{\parallel} \sigma_i^z,
\label{eq:hamiltonian:environment}
\end{equation}
where $\hat \delta(t)=\sum_{n=-\infty}^{\infty} \delta(t-n)$ and $\vec{\sigma}_N \equiv \vec{\sigma}_0$.
The first term corresponds to a homogeneous Ising interaction strength; $b^{\perp}$
and $b^{\parallel}$ are the perpendicular and parallel components of the
magnetic field with respect to the direction of the Ising interaction; finally,
$\hat \delta(t)$ is a train of Dirac $\delta$s with period $1$.
This system has three well-known dynamical regimes.
For both $b^{\perp}=0$ or $b^{\parallel}=0$ the chain is integrable
\cite{prosen}. For
$b^{\parallel}=b^{\perp}\approx \sqrt{2}$ the dynamics is chaotic in the sense of
random matrix theory~\cite{prosenycarlos}. It follows that the nearest neighbor
spacing distribution $P(s)$ of the quasienergies resembles
the one of the circular orthogonal ensemble,
see the appendix. The third regime is an
intermediate one where there is level repulsion but the system is 
not fully chaotic. 
The Floquet operator is 
\begin{equation}
U= 
  \exp\left(-i \sum_{i=0}^{N-1} b^{ \perp }\sigma_i^x +b^{\parallel} \sigma_i^z \right)
  \exp\left(-i \sum_{i=0}^{N-1} \sigma^z_i \sigma_{i+1}^z\right),
\label{eq:floquet}
\end{equation}
and the evolution operator for longer times is simply $U(n) = U^n$.  This model
has the advantage that it can be split in one and two qubit operations, as
the terms in each of the exponentials commute with one another, and one can thus
express the exponential as a multiplication of exponentials each with only one
or two particles involved. 

In order to map local features of the non-Markovianity and have initially null
correlations in any part of the complete system, we use the spin coherent
states as initial states of the environment. 
They are invariant under permutations and can be
regarded as a macroscopic state.

Coherent states are defined as a coherent displacement of the fiducial state
$\ket{J=j;m_z=j}$:
\begin{equation}
\ket{\vartheta,\varphi}= e^{-i \varphi S_z}e^{-i \vartheta S_y}\ket{j;j}= {\cal D}^{(j)}_{\vartheta,\varphi} \ket{j;j},
\label{eq:coherent:states}
\end{equation}
where the total spin is given by $j=N/2$, ${\cal D}^{(j)}_{\vartheta,\varphi}$
is the rotation matrix in the subspace of spin $j$. These states form a complete
basis in the symmetric subspace. In fact, one can parametrize these states in a 
Poincaré sphere, and rewrite 
\begin{equation}
\ket{\vartheta,\varphi}=
\left( \cos \frac{\vartheta}{2} \ket{0}+\sin \frac{\vartheta}{2} e^{i \varphi} \ket{1} \right)^{\otimes N}.
\end{equation}

The environmental Hamiltonian is invariant under external rotations: The translation 
operator, which takes state $\otimes_i |\psi_i\>$ to state $\otimes_i
|\psi_{i+1}\>$, commutes with \eref{eq:floquet}. This symmetry foliates the 
Hilbert space in quasi-momentum $k$ subspaces \cite{prosenycarlos}. As the translation
symmetry leaves \eref{eq:coherent:states} invariant, such states live in the 
$k=0$ subspaces, and as the evolution respects the symmetry, it will remain in 
such subspace. The calculation of the IPR is thus simply 
\begin{equation}
\text{P}^{-1}_{\vartheta,\varphi}= \sum_{i=1}^{\dim \mcH_{k=0}} \left| \langle \phi_i^{(k=0)} \vert \vartheta, \varphi \rangle \right|^4.
\end{equation}

\subsection{Interaction operator} 
\label{sec:model:interaction}

We shall study three kinds of couplings (local, global and generic), and look 
for common trends and differences. Local and generic couplings will break
the symmetry of the environment, whereas the global one is chosen to maintain
it.  We continue by presenting the local perturbations. 

As mentioned above, the interaction was chosen to induce a dephasing channel,
for sake of simplicity. The operator $V$ appearing in \eref{eq:ham},
can be seen as a perturbation operator of the environment dynamics [see
\eref{eq:hamsim}]. 
For the case of global perturbations, we probed altering either 
the magnetic field or the Ising interaction between neighbors, which 
correspond to choosing $V$ as 
%
\begin{equation}
\field \equiv \delta_1(t)\sum_{i=0}^{N-1} \sigma^x_i,\quad
\ising \equiv \sum_{i=0}^{N-1} \sigma_i^z\sigma_{i+1}^z. 
\label{eq:coupling:global}
\end{equation}
Analogously, for the local interaction of the qubit with the environment, 
we chose the coupling as 
\begin{equation}
\localising \equiv  \sigma_0^z\sigma_{1}^z,\quad \localfield \equiv \delta_1(t) \sigma_0^x,
\label{eq:coupling:local}
\end{equation}
where only two and one qubits of the environment, respectively, interact directly with the central qubit.
Finally, to study the generic case, we consider the simplest choice, inspired 
in ergodicity arguments of quantum chaos~\cite{haakebook}. We 
select $V$ from one of the classical ensembles, namely the 
Gaussian unitary ensemble (GUE). We shall denote that case as $V_{\text{GUE}}$, and 
it corresponds to a global and
structureless perturbation.
\section{Results} 
\label{sec:results}


The unitary dynamics in qubit plus environment [defined by Eqs.~(\ref{eq:ham})
and (\ref{eq:hamiltonian:environment}) and the interactions discussed in
\sref{sec:model:interaction}] induce a specific dephasing channel
\eref{eq:channel} once the initial state of the environment is
specified. In our case, such state is a coherent state \eref{eq:coherent:states},
specified by the parameters $\vartheta$ and $\varphi$.
The environment, a spin chain, will be used in integrable, mixed and 
chaotic regimes, varying $b^{\perp}=0.1$, $1$ and $1.4$
respectively while fixing $b^{\parallel}=1.4$. We use
$b^{\perp}=0.1$ instead of  $0$ for integrable dynamics, 
in order to avoid degeneracies in the spectrum and have a well defined \ipr{}.
Corresponding spectral statistics are presented in the appendix. For all
calculations, we chose the coupling parameter $\epsilon=0.1$.

We performed numerical calculations of the measures of \nm{} using time cutoffs
of $t_{\text{cut}}=10^4$ and a mesh in coherent state parameters
$(\vartheta,\varphi)$ of $\Delta \vartheta=\Delta\varphi=0.1$; the two measures
Eqs.~(\ref{eq:rhp}) and \eref{eq:blp} were slightly modified to accommodate to the
intrinsic discrete time structure of \eref{eq:hamiltonian:environment}. We also
considered a time cutoff in the integrals of the measures, as the fluctuations
caused by a finite dimensional environment would send the aforementioned
measures to infinity.
The \ipr{} of the initial environmental states were calculated
with respect to the eigenbasis of $U^+$ for simplicity. Since we are taking a
small $\epsilon$, the \ipr{} does not vary considerably if instead of 
$U^+$, we consider $U^{-}$ or
a Floquet operator with an intermediate $\epsilon$.

We discuss first the relation of the different measures of \nm{} with respect to the
\ipr{}. Next we study the dependence of these quantities with respect to the
choice of the state of the environment; that is, we study the structure of the
environment that can be seen, studying the decoherence of the qubit. The
section is closed with some comments on the generality of the results when one
varies the dimension of the environment and the total evolution time
considered.

\subsection{Dependence of non-Markovianity on the state localization} 
\label{subsec:versus}
We study the behavior of \nm{}, using $\mcN_{\text{RHP}}$ and
$\mcN_{\text{BLP}}$ in \sref{subsec:versus:BLPandRHP} and
then using 
$\mcN_{\mcK}^{\max}$ and $\mcN_{\mcK}^{\<\cdot \>}$ in
\sref{subsec:versus:our:measures}, with $\mcK$ being $\mcD$ or $\mcG$.
In the first section, we focus in the cases which the coupling is via global
and local nearest neighbor Ising interaction, $\ising{}$ and $\localising{}$
respectively; and a global $V_\text{GUE}$ operator. In the second section, we
focus only on global $\ising{}$ and $V_{\text{GUE}}$. These interactions
represent well what happens for the other cases for each study.

\subsubsection{Using BLP and RHP measures} 
\label{subsec:versus:BLPandRHP}
\begin{figure} 
\centering
\includegraphics[scale=1.0]{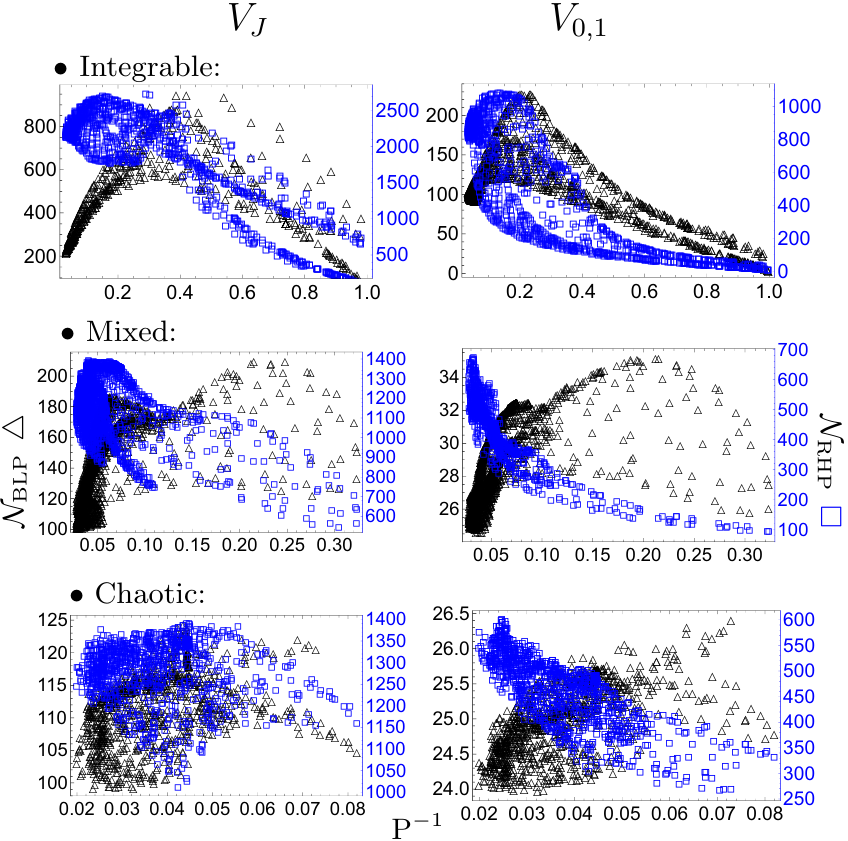}
\caption{BLP (black triangles, left axis) and RHP (blue squares, right axis)
measures as a function of the \ipr{} for initial coherent states of the
environment, \eref{eq:coherent:states}, distributed uniformly on the Poincaré
sphere. Each column corresponds to a different kind of coupling of the
qubit to the environment [see eqs.~(\ref{eq:coupling:global}) and
(\ref{eq:coupling:local})], whereas different rows correspond to different
dynamical regimes of the environment. The parameters used for this and the rest
of the figures are indicated at the beginning of \sref{sec:results}. The results for the global and local field perturbation, $\field{}$ and $\localfield{}$ respectively, are very similar to their global and local Ising counterparts.
\label{fig:rhpvbsblpising}}
\end{figure} 
In \fref{fig:rhpvbsblpising} we show, for different initial conditions of the
environment and a coupling of the type $\ising{}$, the value of \nm{} using BLP
and RHP measures as a function of the \ipr{}.

\begin{figure} 
\centering
\includegraphics[scale=1]{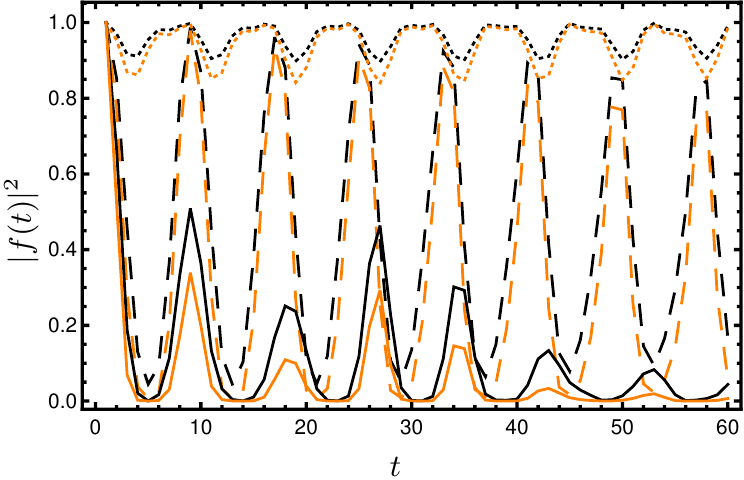}
\caption{Typical behavior of the fidelities of the environment,
\eref{eq:hamiltonian:environment}, in the integrable regime with 
a global Ising perturbation  $\ising{}$, for several
coherent states \eref{eq:coherent:states}.  We consider $10$ and $16$ qubits,
shown in black and orange curves respectively. The figure shows the fidelity
for the state $\ket{\vartheta=2.8, \varphi=4.8}$ (with
IPR equal to 0.457 and 0.375 for $10$ and $16$ qubits respectively)
which is among the states that yield larger values for measures based on
$\mcD(t)$ (dashed curves). Fidelities for the states that give low values of
the BLP measure are the dotted and solid curves, obtained from the states $\ket{\vartheta=3.0,
\varphi=2.2}$ (IPR equal to 0.994, 0.984) and $\ket{\vartheta=1.5,
\varphi=3.5}$ (IPR equal to 0.046, 0.010), respectively, which are high and low localized states.\label{fig:fidelities}
}
\end{figure}

%
In the integrable regime the two measures have different behaviors;
$\mcN_{\text{BLP}}$ grows for increasing IPR until it reaches a maximum around
$\text{P}^{-1} \sim 0.4$, where it starts to decrease.  $\mcN_{\text{RHP}}$ has
an approximate monotonic decreasing behavior, showing a change of slope around
$\text{P}^{-1}\sim 0.4$ and another close to $\text{P}^{-1}\sim 0.6$.  A local coupling, namely $\localising$, yields similar results;
however, the peak in the BLP measure is sharper and the decay of RHP measure is
faster (\fref{fig:rhpvbsblpising} second column).
%
The behavior of $\mcN_{\text{BLP}}$ can be explained qualitatively by studying the
fidelity which, for the dephasing case, is related to the distinguishability
via the equation
$\mcD(t)=\vert f(t) \vert^2$.  In \fref{fig:fidelities} we show its evolution in
the integrable regime, for three initial conditions and two different
environment sizes.  For high and low values of localization, oscillations of
$\mcD(t)$ are constrained around high and low values of asymptotic fidelity,
respectively. Therefore, the relatively low values of non-Markovianity belong
to the high and low values of localization. There are also states with
high \ipr{} that lead to distinguishabilities that oscillate with large
amplitude but at a low frequency; those states have low asymptotic fidelity.
The maximum value of  \nm{} is achieved at $\sim 0.4$, where fidelity can
oscillate with a large amplitude.  One can understand the behavior of
$\mcN_{\text{RHP}}$ with similar arguments [see \eref{eq:rhppractical}] but
this time taking into account the role of the logarithm. For high localized
states the typical values of the minimums and maximums of $\mcD(t)$ are very close
to one or with lower frequency, yielding very small values
of the logarithm and thus
low values of the RHP measure. As the \ipr{} decreases, the minimums in
$\mcD(t)$ diminishes faster than the maximums, and one reaches quickly the
regime in which 
$-\log(F(a_i))\sim
{\mathcal{O}}(1)$, causing an increasing of the measure until
$\text{P}^{-1}\sim 0.4$.
For small values of the localization,
the typical minimum is very close to zero, 
for which the logarithm is large, in absolute value. One can approximate
$\mcN_{\text{RHP}}\approx \sum_i
\log(F(b_i))+n \log(F(\tilde a^{-1}))$, where $n$ is the number of minimums included in
the interval of the computation of the measure and $F(\tilde a)$ is its typical
value. The value of the measure is now seen to be directly related
with the localization giving again a monotonic behavior with different slope.
For both measures, low localized states tend to cluster. These states are 
localized in the equator of the Poincaré sphere (see
\fref{fig:densityising}). This explains the two leaf-like structures
connected by a stem in the integrable regime. 
\par

\begin{figure} 
\centering
\includegraphics[scale=1]{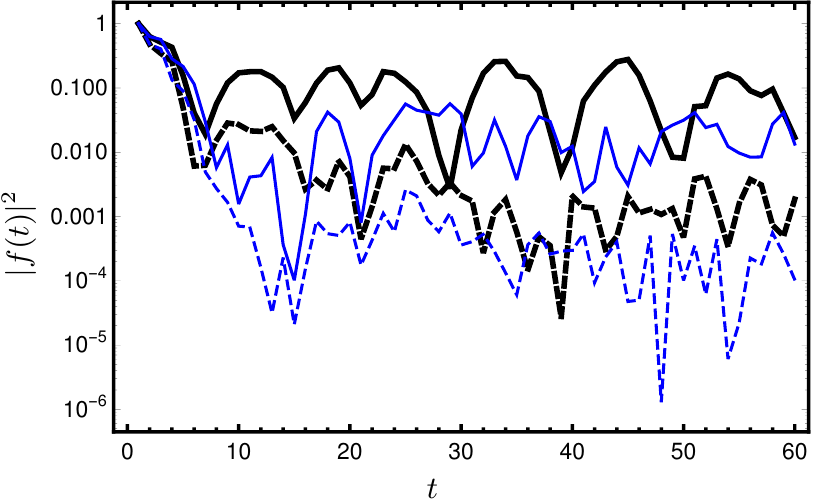}
\caption{Typical behavior of the fidelities of the environment in the chaotic
(blue curves) and mixed (black thick curves) regimes for $10$ (solid curves) and $16$
(dashed curves) qubits, with the coupling $V=\ising{}$; the initial state is a coherent state characterized by 
$\ket{\vartheta=0.7,\varphi=0.8}$, see \eref{eq:coherent:states}.
A fast decay and fluctuations around a value determined by the effective
dimension of the Hilbert spaces, explains the values of the different
measures of non-Markovianity. 
\label{fig:fidelitiescao}}
\end{figure}

In the mixed and chaotic regimes fidelities begin with a fast decay after which
they fluctuate around the inverse of the effective dimension of the state
(\fref{fig:fidelitiescao}). Since the asymptotic fidelity is inversely
proportional to the effective dimension of Hilbert space, the scale of the
\nm{} is lower in these regimes with respect to the integrable. 
The \ipr{} is also small due to ergodic properties of the Hamiltonian.
In the mixed regime the slope of the data using $\ising{}$ and $\localising{}$
is positive for BLP measure, while for RHP it is clearly
decreasing for both perturbations, mimicking the integrable cases.  Thus, 
both measures behave
differently also in the mixed regime.
In the chaotic regime, we expect full ergodic properties, and consequently, 
a similar reasoning to that of the mixed case will follow, however with smaller
IPR. Indeed, all initial conditions cluster around a smaller region but
a slope, consistent with the mixed cases, is observed.
\par

\begin{figure} 
\centering
\includegraphics[scale=1]{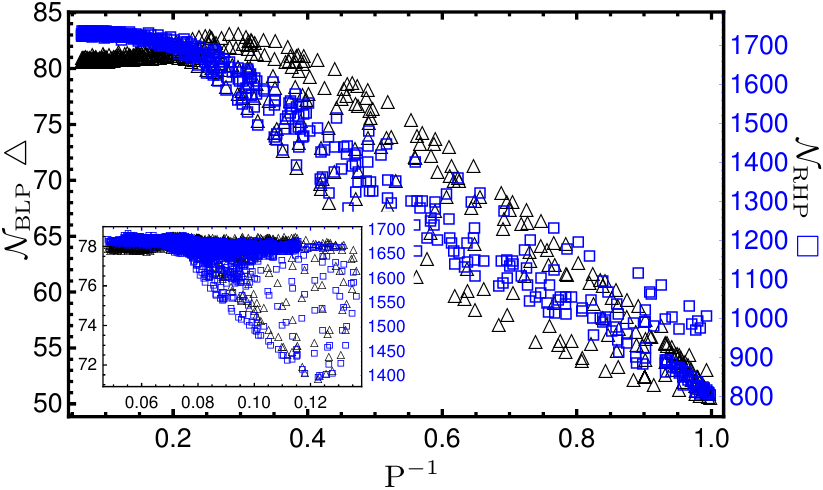}
\caption{RHP and BLP measures of the spin chain using a random coupling, chosen 
from the GUE, in the
integrable regime (main panel) and  the chaotic regime (inset). We observe a
monotonic decreasing behavior for both measures in all regimes, with a short
growth for BLP measure in the integrable regime.\label{fig:blprhprmt}}
\end{figure}

Finally, we show the results when a random potential provides the coupling 
in \eref{eq:ham}; namely, when we take $V=V_\text{GUE}$. 
The dependence of \nm{} on the \ipr{} is shown in
\fref{fig:blprhprmt} for both the integrable and the chaotic cases.
Its behavior is qualitatively similar to the one observed for the other
couplings, when comparing among integrable cases, mixed and chaotic ones. 
However, there are some quantitative differences. For example, 
the BLP measure still has an initial growth but is very short compared
with the case of $V=\ising{}$. The same arguments as before
can be stated to explain the general features of the behavior. 
%
%
\subsubsection{Using measure schemes $\mcN_{\mcK}^{\max}$ and $\mcN_{\mcK}^{\langle \cdot \rangle}$} 
\label{subsec:versus:our:measures}
In the previous section we considered measures BLP and RHP, which are based on 
the non monotonicity of distinguishability, as measured by $\mcD(t)$,
and of divisibility, as measured by $\mcG(t)={\int_0^t
g(\tau) d\tau}$.
In this section we use measures based on the same
quantities, but use Eqs.~(\ref{eq:GorinNM}) and~(\ref{eq:CarlosNM}) to
obtain a quantity that can be directly related to a physical
process~\cite{ourmeasure}, and contrast its behavior with 
measures BLP and RHP. 

\begin{figure} 
\centering
\includegraphics[scale=1]{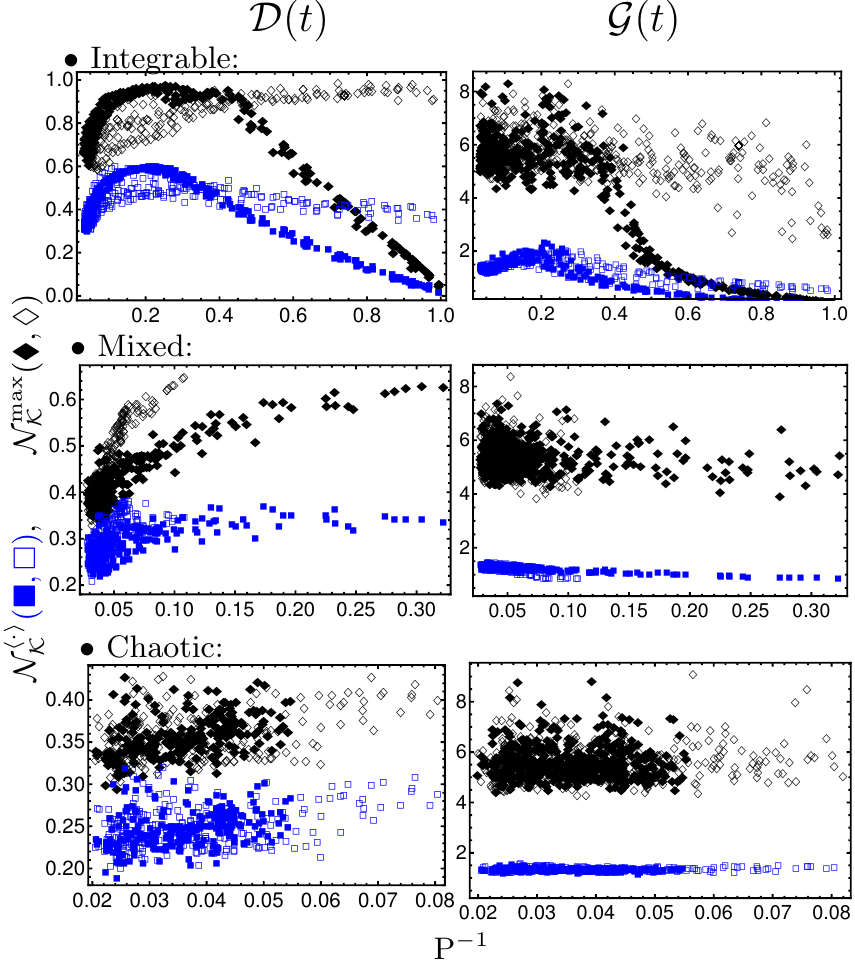}
\caption{Measures $\mcN_{\mcK}^{\langle \cdot \rangle}$ (blue) and $\mcN_{\mcK}^{\max}$
(black) with $\mcD(t)$ [left column] and $\mcG(t)$ [right column], for the spin
chain using global Ising perturbation $\ising{}$, as a function 
of the initial \ipr{} of the environment, see \fref{fig:rhpvbsblpising}. 
The initial states of the environment are coherent states uniformly 
chosen from the northern/southern hemisphere of the Poincaré sphere and 
indicated by the hollow and filled markers, respectively.
In the integrable regime (and in the mixed for $\mcN_{\mcD}^{\max}$) 
we see two different behaviors, coming from the two hemispheres of the Poincaré
sphere. The results for local Ising interaction, $\localising{}$, are very similar to the presented here. Results for global and local field perturbations, $\field{}$ and $\localfield{}$ respectively, presented only the behavior plotted by filled markers.  
\label{fig:ourmeasureising}}
\end{figure}

For the integrable case, we observe that there are two distinct behaviors, 
for both measures $\mcN^{\max}_{\mcK}$ and $\mcN^{\langle \cdot \rangle}_{\mcK}$, 
regardless of whether they are based on $\mcD$ or $\mcG(t)$. 
In \fref{fig:ourmeasureising}, we show the results for the case in which
the coupling is $\ising{}$. These two different behaviors are associated with the two
hemispheres of the 
Poincaré sphere, and its details can be understood by studying the 
evolution of fidelity. 
In particular, for $\mcN^{\max}_{\mcD}$, one of the branches
displays a maximum ($\text{P}^{-1} \sim 0.4$), then it decays
linearly. The other branch, corresponding to the southern hemisphere ($\pi/2 <
\vartheta \le \pi$), has a
slight increase with IPR. 
The behavior of $\mcN^{\langle \cdot \rangle}_{\mcD}$ is similar;
however, it is scaled down, and instead of a slight increase, 
the southern hemisphere displays a small increase with IPR. 
A quantitatively similar behavior is seen when we 
 base our measures in $\mcG(t)$, with the bending
point  being again at 
$\text{P}^{-1} \sim 0.4$, for $\mcN^{\max}_{\mcG}$.
$\mcN_{\mcG}^{\langle \cdot \rangle}$ 
is also a scaled down and slightly deformed version 
of $\mcN_{\mcD}^{\langle \cdot \rangle}$.
%
For low localized states, the explanation of the aforementioned behavior
is similar to the one given for BLP and RHP measures. 
Since the size of the fluctuations of the fidelity depend on the
effective dimension of the state, 
 $\mcN^{\langle \cdot \rangle}_{\mcK}$ and
$\mcN^{\max}_{\mcK}$ increase as we take more
localized initial environmental states. For highly localized states in the
integrable regime, there are two families of states. One, with 
asymptotic fidelity greater than
$1/2$ and whose fidelity has a high frequency, but small amplitude, and 
other with asymptotic fidelity smaller than $1/2$ but  with a fidelity that has
smaller frequency and a larger oscillation amplitude. 
Since the schemes under
discussion depend mainly in the amplitude of the oscillations, they are
critically sensitive to the asymptotic fidelity of the environmental states.
This feature is a significant difference between the newly proposed
schemes~\cite{ourmeasure} and the more often used BLP and RHP. 


In the mixed and chaotic regimes, the behavior of the measures is monotonically 
increasing. Since all coherent states have a small \ipr{}, the same
arguments given before for low localized states  in the integrable 
regime hold to explain such monotonicity. For the chaotic regime the measures
also tend to homogenize; this
is expected given that the initial states have similar effective dimension, 
as they appear random in the eigenbasis of the Floquet operator.

\begin{figure}
\centering
\includegraphics[scale=1]{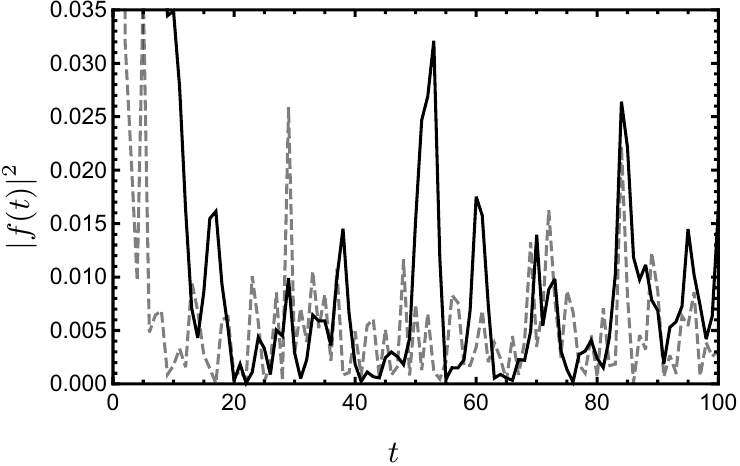}
\caption{Typical behavior of the fidelities in the integrable regime for
$V=V_{\text{GUE}}$. In solid black we plot the fidelity of the state
$\ket{\vartheta=3.2,\varphi=1.1}$ as a representative state of highly localized
states, and $\ket{\vartheta=2.2,\varphi=2.4}$ in dashed gray as a
representative of low localized states. 
High localized states lead to a low frequency of occurrence of pairs of local
minima and maxima, while for localized states such frequency is increased.
This explains the different behaviors among \blp{} and $\mcN_{\mcD}^{\max
,\langle \cdot \rangle}$. 
\label{fig:fidelities_rmt}
}
\end{figure}

\begin{figure}
\centering
\includegraphics[scale=1]{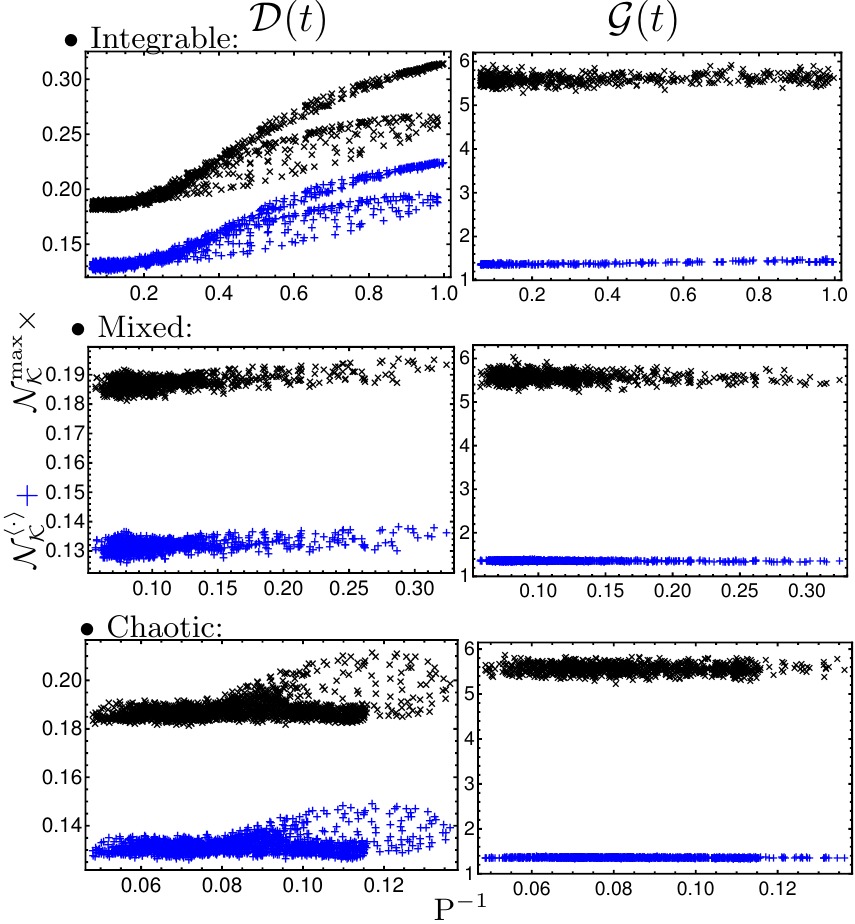}
\caption{Relation between 
$\mcN_{\mcK}^{\langle \cdot \rangle}$ and $\mcN_{\mcK}^{\max}$ with \ipr{}, 
using a global random perturbation. 
We consider the two measures, based on both 
$\mcD(t)$ and $\mcG(t)$ and a spin chain of eight spins for an 
ensemble of $40$ matrices. Measures based on $\mcG$ are
almost constant in all regimes. For 
 $\mcN_{\mcD}^{\max, \langle \cdot \rangle}$ in the integrable regime, we observe 
different behaviors for each hemisphere of the Poincaré sphere. 
\label{fig:ourmeasurermt}}
\end{figure}

For a random coupling to the environment, measures 
$\mcN_{\mcD}^{\max, \langle \cdot \rangle}$ have a monotonic behavior with 
respect to \ipr{}. However, in contrast to the behavior of 
the BLP measure, non-Markovianity increases with the inverse participation
ratio. 
This surprising change can be explained when noticing that
the BLP measure depends on the number of pairs of minima and
maxima that appear in the fidelity in a given interval, while 
$\mcN_{\mcD}^{\max, \langle \cdot \rangle}$ depend {\it only} on the amplitude of the
fluctuations of $F(t)$. As we take more localized initial
environmental states, the size of the fluctuations is increased as the pairs of
minima and maxima appear less frequently (shown in \fref{fig:fidelities_rmt}), 
which explains the aforementioned effect.
The behavior in the mixed regime, which is also monotonic increasing, has the
same explanation. In the chaotic regime the values of \nm{} also tend to
homogenize, having the same explanation as the one given for $V=\ising{}$ for
this regime. 
Now using $\mcG(t)$ as indicator, all measure
schemes in all regimes yield almost constant \nm{} with respect to the \ipr{}
(right panels
of \fref{fig:ourmeasurermt}). This
behavior is expected for the chaotic regime; what remains to be explained is its
emergence in the integrable and mixed regimes. 
To do this we can find an upper bound for the change of $\mcN_{\mcG}^{\max}$ in the whole interval of localization; we shall call this
$\Delta\mcN_{\mcG}$. 
From \eref{eq:GorinNM}, $\mcN_{\mcG}^{\max}=\log\left(F(t_f)\right)-\log\left(F(\tau)\right)$, where $t_f$
and $\tau$ are the maximum and the minimum attained to the maximization
required by the definition. Now, since the logarithm is a monotonic function,
the measure $\mcN_{\mcD}^{\max}$ is attained to the same times, allowing us to
write $\mcN_{\mcG}^{\max}=\log \left(
\mcN_{\mcD}^{\max}+F(\tau)\right)-\log \left( F(\tau) \right)\approx \log
\left( \mcN_{\mcD}^{\max} \right)+F(\tau)/\mcN^{\max}_{\mcD}-\log(F(\tau))$.
Therefore the total change is $\Delta\mcN_{\mcG}=\Delta \log\left(
\mcN_{\mcD}^{\max}\right)+\Delta \left(
F(\tau)/\mcN_{\mcD}^{\max}\right)-\Delta \log \left(F(\tau)\right)$. The last
term can be ignored since $F(\tau)$ is typically very similar for any value of
localization. The second term is negative since $\mcN_{\mcD}^{\max}$ changes
faster than $F(\tau)$ and its absolute value is smaller than the first term
which is positive. 
Therefore $\Delta \mcN_{\mcG}^{\max}$ is upper bounded by $\Delta \log\left(
\mcN_{\mcD}^{\max}\right)$ and its numerical values for the integrable and mixed regime are $0.4$ and $0.08$ respectively. There is a similar explanation
for $\mcN^{\langle \cdot \rangle}_{\mcG}$ using typical values of the
average instead of the minima.

We finish this section by summarizing the results and commenting on practical
consequences of the relations we found between non-Markovianity and IPR.
The integrable regime shows the richest behavior when we use a structured
coupling to the environment. In our case we observed a wide variety which includes
up to two different behaviors for the two hemispheres of the Poincaré sphere. 
In general the different measures behave differently and depend on the details 
of the fidelity. However, the IPR determines coarsely the value of the 
non-Markovianity. 
As mentioned in \sref{sec:measures}, measure $\mcN^{\langle \cdot \rangle}_{\mcD}$
is directly related to the task of storing information safely; we 
can see that to perform such a task with a high probability of success, we need 
an environment in the integrable regime, a structured interaction, and states
with intermediate localization.  When the environment is in the chaotic regime, the behavior is not so rich, as the coherent states are quite delocalized, and the non-Markovianity seems to be self
averaging. In the mixed regime of the environment, we have an intermediate
behavior. \par
\subsection{Underlying structure} 
In Ref.~\cite{mata2}, the authors show that non-Markovianity, via long time
fluctuations of fidelity, is able to resolve complex phase space structures of
the environment using initial coherent states. In particular, the fractal
nature of the phase space is clearly visible in the mixed regime.  We
investigated the spin chain in a similar way, using spin coherent states as
initial environmental states, studying now the measures of \nm{} and the \ipr{}
as functions of the parameters of the spin coherent states. Our goal is to
study the visible structures and how they change during the transition from
integrability to chaos. \par
\begin{figure*} 
\centering
\includegraphics[scale=1]{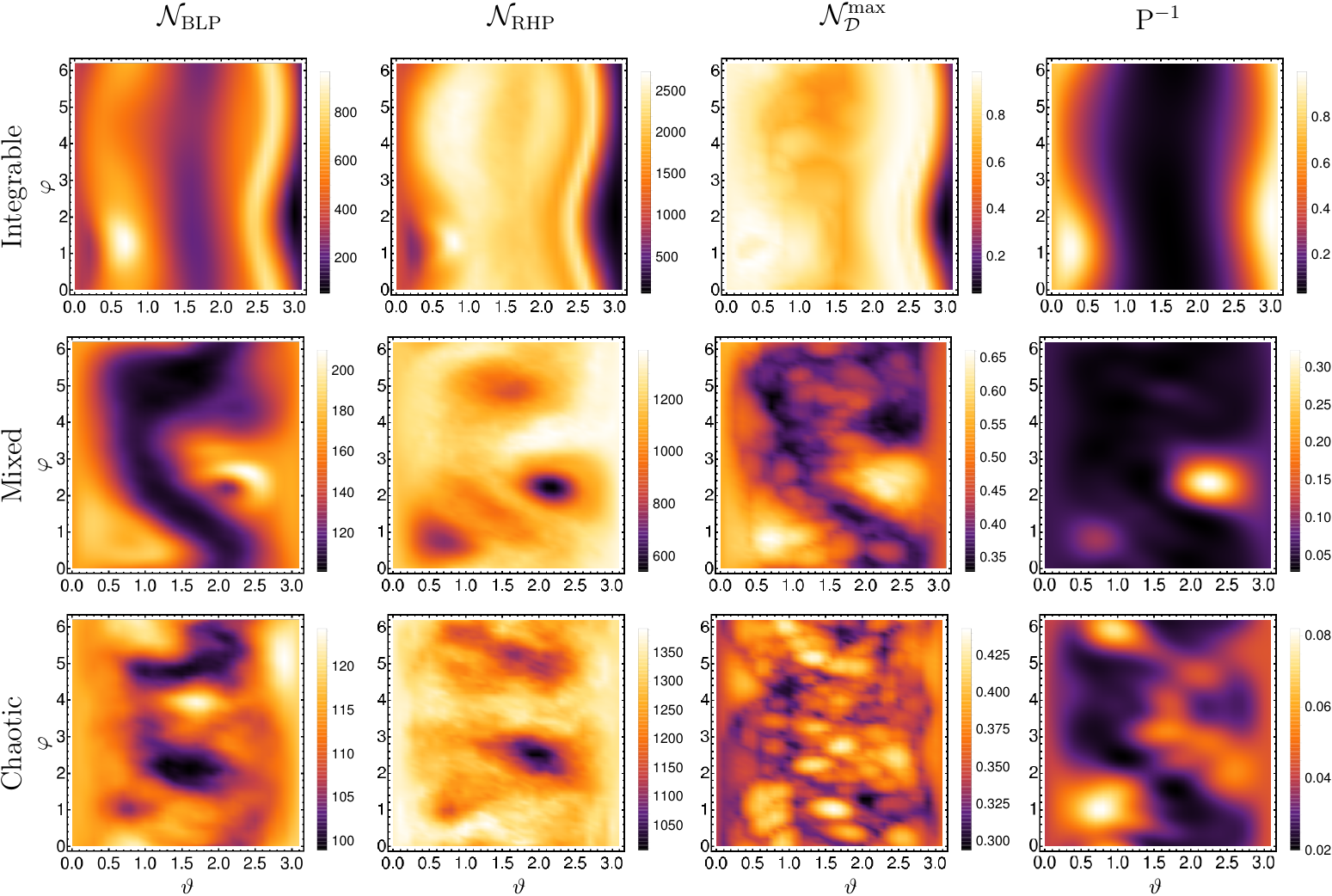}
\caption{The different columns correspond to
density plots of several measures of non-Markovianity and the IPR,
for a chain with $10$ qubits using the homogeneous perturbation 
$\ising{ }$. 
For  $\mcN_{\mcD}^{\max}$ some smaller structures appear as we go into 
the chaotic regime. We can also observe a relation in the 
integrable  and mixed regimes between \ipr{} and all  non-Markovianity measures. The results for local Ising interaction, $\localising{}$, are very similar, just with more extended depressions . The results for the global and local field perturbations, $\field{}$ and $\localfield{}$ respectively, are very similar to their global and local Ising counterparts.
\label{fig:densityising}}
\end{figure*}

 
In the integrable regime (top of \fref{fig:densityising}), the values of
the \nm{} measures mimic the behavior of the \ipr{} close to the equator of the
Poincaré sphere ($\vartheta=\pi/2$); 
 close to the poles the situation is different.
%
The equator of the Poincaré sphere corresponds to low localized states, and
 this in
turn leads to local minimums for all examined measures of \nm{}.
When moving toward the poles, which are very localized states, one finds a
local maximum, and then in the vicinity of the pole, a local minimum, for all
cases except for $\mcN_{\mcD}^{\max}$ near the north pole. 
The difference arises from the different asymptotic fidelities of the
chosen high localized states. 
This picture deepens the understanding of the behavior already seen
in Figs.~\ref{fig:rhpvbsblpising} and \ref{fig:ourmeasureising}.\par
For the mixed regime, the features on the \nm{} measures are mainly governed by
the \ipr{}. High localization leads to local maximums in the
measure $\mcN_{\mcD}^{\max}$ and local minimums for the \rhp{} measure. For the \blp{}
measure, there is also an interesting feature. The local maximum of \ipr{},
located around $\vartheta\approx\varphi\approx 2.5$, leads to a local
minimum on the \nm{} which is partially surrounded by a maximum. This
behavior is actually similar to the one at
the poles in the integrable regime.  In the chaotic regime
the relation of the measures with the localization practically vanishes.
\par 
Regarding the transition from integrability to chaos, using the BLP and RHP
measures, there is not a notable change in the size of the structures 
as it does for environments with a classical analog~\cite{mata,mata2}.
This might be due to the absence of such structures, or, that simply due to the
relative size of the coherent states in this system, they are 
not able to resolve small structures. 
%
%
More quantitatively, the fluctuations of the
spin coherent states in the Poincaré sphere (chosen to have radius one) scale
as $\sim N^{-1}$~\cite{klimovbook}
\ie{} as $\left[\log_2 \left( \dim \mcH
\right)\right]^{-1}$, while for coherent states in the torus 
fluctuations scale as $(\dim \mcH)^{-1}$~\cite{Garcia-Mata2012}.\par
The situation is different when using $\mcN^{\max}_{\mcD}$. In the transition
to chaos, a finer structure emerges.  Although such features do not appear
classical, in the sense of the appearance and breaking of KAM tori, it is
clear that there is a finer granularity than is typically expected in this
transition; these structures are robust with respect to changes in parameters and 
times of integration. We consider this one of the central results of this work. \par

\begin{figure*} 
\centering
\includegraphics{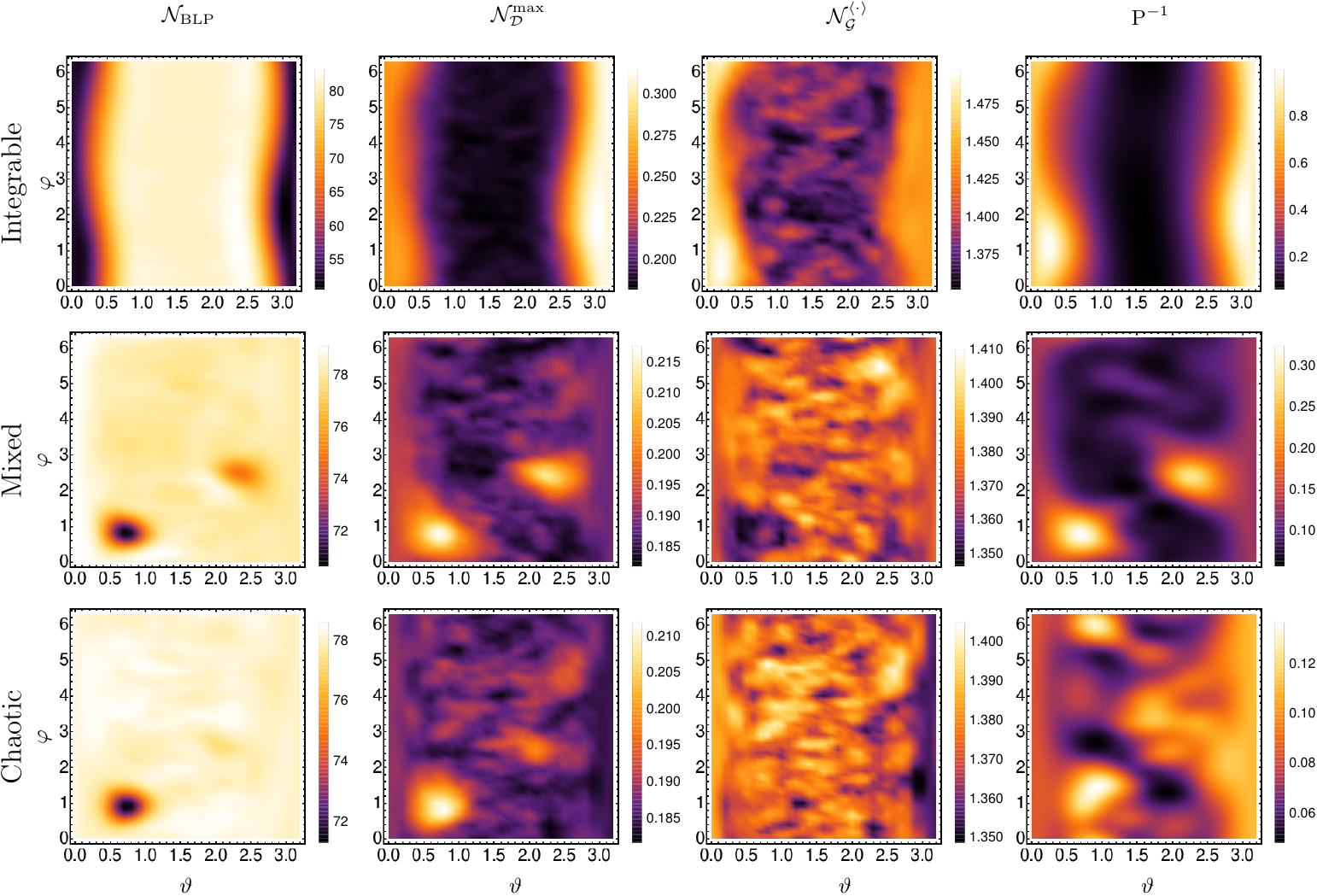}
\caption{Density plots of the \nm{} measures and the IPR for the
chain with $8$ qubits, using random potentials $V=V_{\text{GUE}}$ averaged over
$40$ matrices. Figures show an emerging fine structure in the transition from
integrability to chaos in $\mcN_{\mcG}^{\langle \cdot \rangle}$. The
structures observed in the \nm{} measures are correlated (or anticorrelated)
with the \ipr{}. \label{fig:densityrmt}}
\end{figure*}

Let us now comment on the results using $V=V_{\text{RMT}}$, shown in~\fref{fig:densityrmt}.
In the integrable regime, measures $\mcN^{\max}_{\mcD}$ and
$\mcN^{\langle \cdot \rangle}_{\mcG}$ completely mimic the behavior of
the \ipr{}, while the \blp{} measure is anticorrelated with the \ipr{}. Such
behavior is a consequence of the way fidelity contributes 
to the different measures. 
Recall that the \blp{} measure depends mainly on the
frequency with which the pairs of minima and maxima occur in $\mcD(t)$,
while schemes $\mcN_{\mcK}^{\max}$ and $\mcN_{\mcK}^{\langle \cdot
\rangle}$ depend mainly on the amplitude.
In the mixed and chaotic regimes the situation is similar:
The \ipr{} is correlated with $\mcN^{\max}_{\mcD}$ and anticorrelated
with  \blp{}. However, for 
$\mcN^{\langle \cdot \rangle}_{\mcG}$ the landscapes appear
to have almost no correlation with IPR. 

It is also important to underline that the measures $\mcN_{\mcG}^{\langle
\cdot \rangle}$ and $\mcN_{\mcD}^{\max}$ show a non-fractal structure in
the transition to chaos, as in the results using $\ising$.

%
Results using RHP measure are very similar to the ones for BLP; the ones for
$\mcN_{\mcD}^{\langle \cdot \rangle}$ resemble the ones for $\mcN_{\mcD}^{\max}$,
and the results using $\mcN^{\max}_{\mcG}$ reveal only a random
landscape for all regimes. 

\begin{figure} 
\centering
\includegraphics[scale=1]{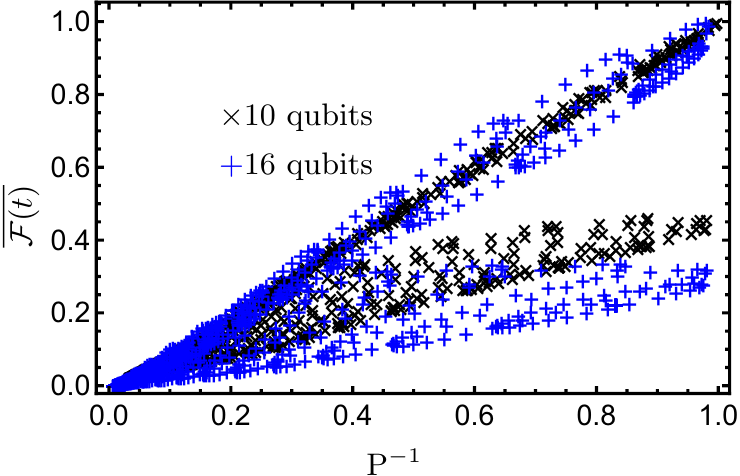}
\caption{Relation of the \ipr{} and the averaged square root of the asymptotic fidelity, $\overline{\mcF(t)}$, for
$10$ and $16$ qubits. The figure shows the splitting in
the \nm{} vs \ipr{} relation, explaining both the peculiar behavior of the
results shown in  \fref{fig:ourmeasureising} and the spreading of the non-Markovianity
measures, for a fixed IPR, as the dimension is increased.
\label{fig:allisingfidipr}}
\end{figure} 
%
\subsection{Generality of the results} 

This section is devoted to a discussion the validity of the main results
presented above
for a larger number of qubits and for different cutoff times. 

%

\begin{figure*}[h] 
\centering
\includegraphics[scale=1]{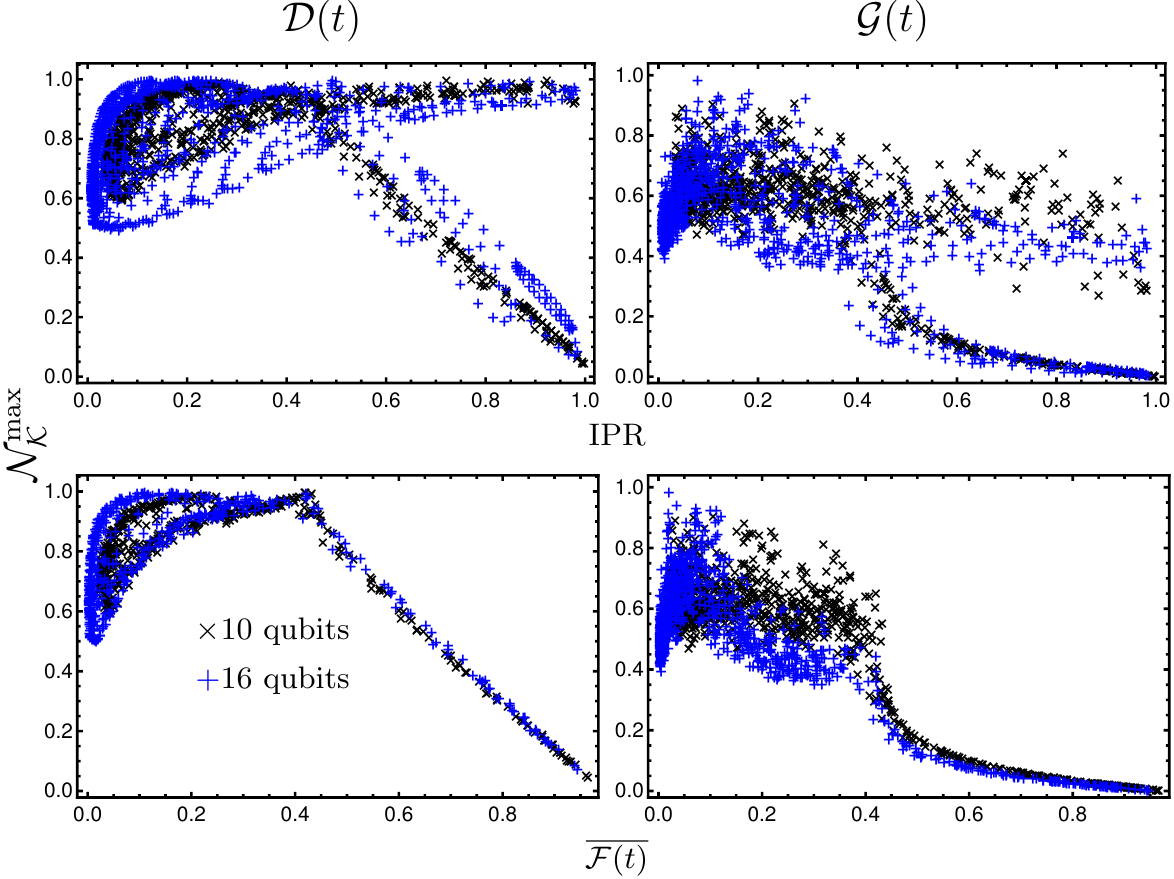}
\caption{Relation of the \nm{} with the \ipr{} and with the averaged asymptotic
fidelity
$\overline{\mcF(t)}$, for $10$ and $16$ qubits using $V=\ising{}$ (compare with
\fref{fig:ourmeasureising}). The figures show the data
spreading of the relation \nm{} versus \ipr{} when the dimension is increased
(upper row).
Such feature is not present in the relation of \nm{} versus
$\overline{\mcF(t)}$ (lower row). We have the same situation 
\label{fig:allisingmaxg}}
\end{figure*}

We first discuss three key features, namely (i) the decreasing behavior of the
\blp{} and \rhp{}
measures for high localized states (shown in Figs. \ref{fig:rhpvbsblpising}
and \ref{fig:blprhprmt}); (ii) the same property for measures
$\mcN_{\mcK}^{\max}$ and
$\mcN_{\mcK}^{\langle \cdot \rangle}$, but only for the hemisphere which contains the
states with low asymptotic fidelity (\fref{fig:ourmeasureising}); and (iii) the peculiar behavior of measures based on $\mcD(t)$ (also shown in \fref{fig:ourmeasureising}), which exhibits a clear change on the slope as localization is increased. Let us now comment how these observations behave as the dimension of the environment is increased, and for sake of brevity only for measures $\mcN_{\mcK}^{\max}$ (shown in~\fref{fig:allisingmaxg}). The results show that the patterns are preserved; however, as the dimension increases the data becomes diffused, \ie{} for each value of \ipr{} there is a wider range of \nm{}.
This is due to the relation between asymptotic fidelity
$\overline{\mcF(t)}$ and \ipr{} (shown in top panel of~\fref{fig:allisingfidipr}), which is
linear (for each hemisphere of the Poincaré sphere) but spreads out for a
larger number of qubits.

It is interesting that this observation also reveals the origin of the above
mentioned splitting of the relation between localization and non-Markovianity,
due to the
different values of asymptotic fidelities of high localized states. Therefore,
by plotting the relation of \nm{} versus $\overline {\mcF(t)}$ (shown in the
bottom panel of~\fref{fig:allisingmaxg}), it can be seen that the splitting
and the spreading of the data are removed, revealing that the relation of \nm{}
is simpler as a function of the effective dimension of the Hilbert space of the
initial states.
%

\begin{figure*} 
\centering
\includegraphics[scale=1]{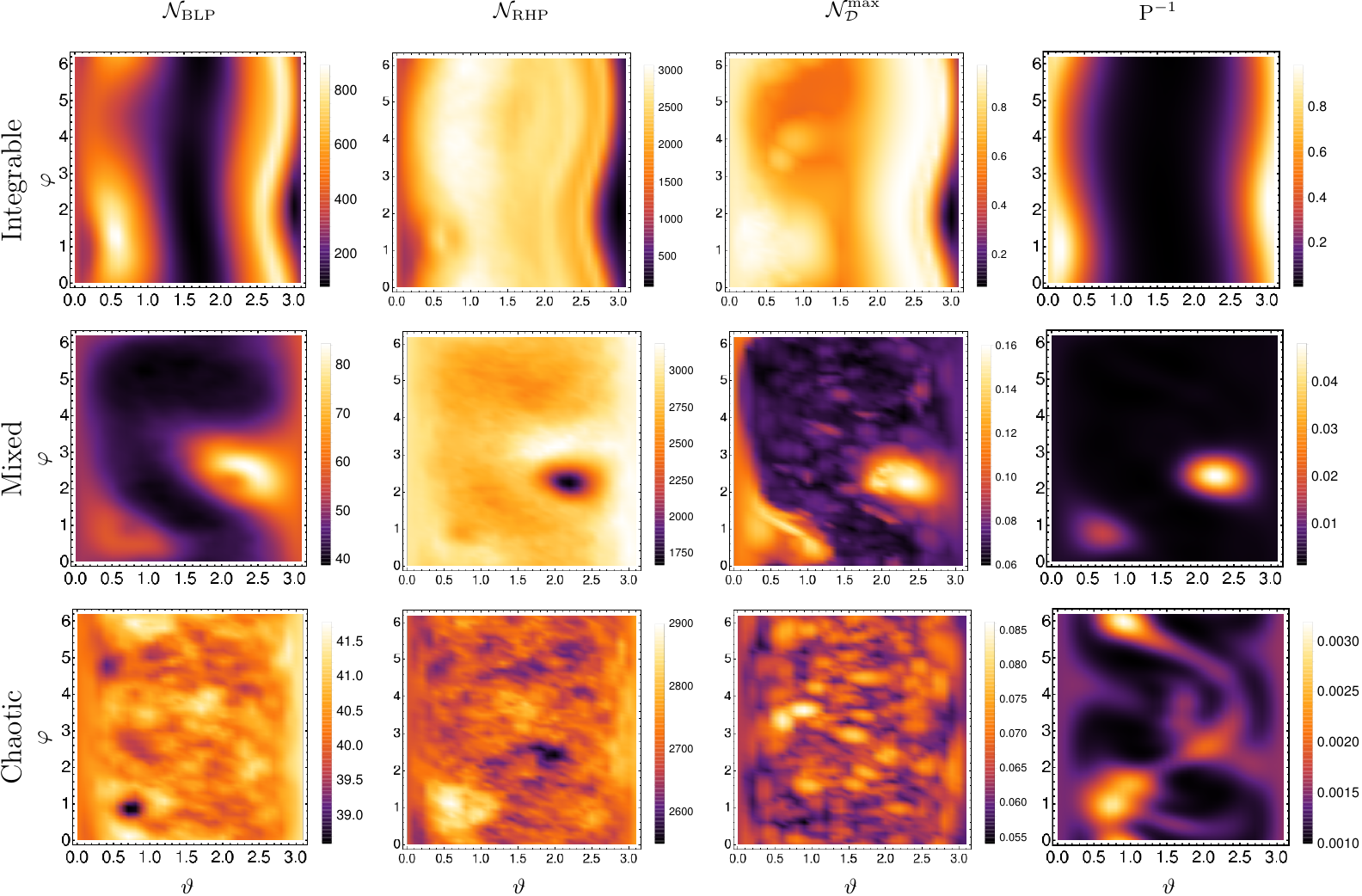}
\caption{
Density plots of the measures of \nm{} and of the \ipr{} for
the chain with $16$ qubits and $V=\ising{}$. As in the case of the chain with
$10$ qubits, the fine structures in $\mcN_{\mcD}^{\max}$ is present. The local
maximums in the \ipr{} also dictate where are the dominant structures of local
maximums or minimums in the \nm{}. \label{fig:density16}}
\end{figure*}

Next, we shall study the emergent structures in the computed measures
for the system with a higher dimension (we used a spin chain with $16$ qubits).
It yields basically the same behavior as for the $10$ qubits case (shown in
\fref{fig:density16}), but there is an emergence of smaller
finer features in the landscapes of measure $\mcN_{\mcD}^{\max}$
($\mcN_{\mcG}^{\langle \cdot \rangle}$), which has basically an identical landscape.
We conclude that such
fine structures become smaller as the dimension is increased.  A general
characteristic of the landscapes, especially in the integrable and mixed
regimes, is that the local maximums in the \ipr{} determines the most visible
structures in the \nm{}. They appear as local maximums or minimums depending on
the chosen measure and/or in the asymptotic averaged fidelity of
the coherent states of the region.
\par
\begin{figure} 
\centering
\includegraphics{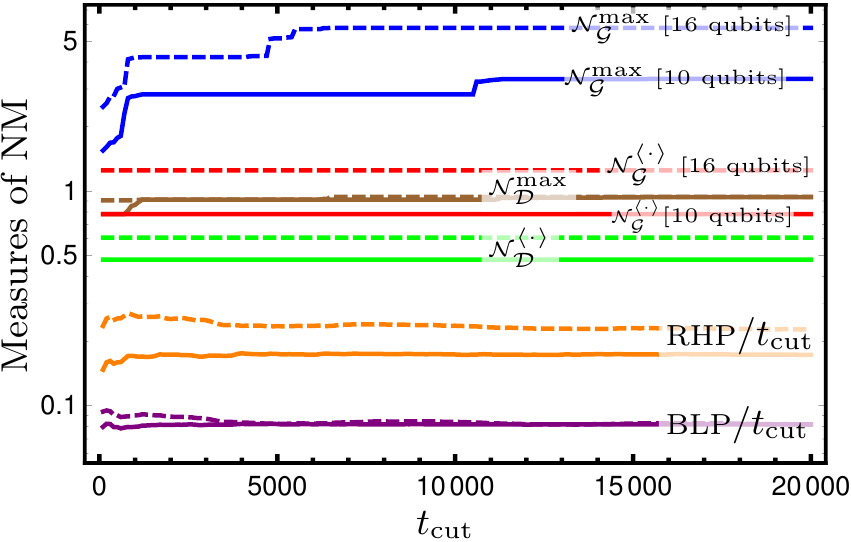}
\caption{Measures of \nm{} as a function of the cutoff time,
using the coherent state $\ket{\vartheta=2.8,\varphi=4.8}$ 
for $10$ (solid lines) and $16$ (dashed
lines) qubits in the integrable regime. \blp{} and \rhp{} measures are
normalized by $t_{\text{cut}}$
to remove their linear dependence on $t_{\text{cut}}$; it is clear from this 
plot, that without such normalization they grow mainly linearly with time.
The figure shows that the
cutoff time used throughout the paper is appropriate to 
understand the results for asymptotic times. }
\label{fig:timestability}
\end{figure}

 
We finalize this section by discussing the validity of our observations for
other cutoff times. In Fig.~\fref{fig:timestability}, we show the values of
all the measures
treated in this paper for the integrable case and for one state of the
environment, as a function of the cutoff time. Measures \blp{} and \rhp{} are
normalized by $t_{\text{cut}}$ to avoid their trivial linear dependence.
The figure shows that all measures saturate
quickly to its asymptotic value, except $\mcN_{\mcG}^{\max}$, which
saturates more slowly than others but  more quickly with respect to the system size. We
discussed only the results for one state in one regime since the exploration
for other cases gives very similar results. \par 

%
%
\section{Conclusions} 
\label{sec:conclusions}
We performed numerical calculations of the non-Markovianity of a qubit coupled
to an environment modeled by a unitary kicked spin chain in a coherent state.
Several dynamical regimes of the chain, couplings between qubit and
environment, and measures of non-Markovianity, were considered. Additionally,
the inverse participation ratio of the environment (with respect to the coupled
environment) was calculated. 

We explored the relation of \nm{} versus \ipr{} and showed that the schemes
$\mcN_{\mcK}^{\max}$ and $\mcN_{\mcK}^{\langle \cdot \rangle}$, proposed in
Ref.~\cite{ourmeasure} have important and potentially useful differences with respect to
the more common measures \blp{} and \rhp{}. We showed that that the first
mentioned schemes reveal the asymptotic fidelity of the environmental state,
leading to two clearly different behaviors of the measures in function of the
\ipr{}.  Regarding the validity of the former results, we showed that the
relations between non-Markovianity and localization for larger environments
remain the same. However, self averaging was not observed.  A central result of
the paper is the identification of a maximum of the \nm{} for intermediately
localized environmental states, when using distinguishability as indicator.
Such a scenario could be used to protect classical
information more efficiently~\cite{ourmeasure}.

In the second part of the work we presented a study of the \nm{} and the \ipr{}
as functions of the parameters of the Poincaré sphere in which 
the initial coherent environmental states live. 
We concluded that there are structures mainly depicted by the \ipr{} in all dynamical
regimes; these are robust under the election of the interaction Hamiltonian and
the dimension of the environment.
We have shown that although such structures are not classical-like (in
the sense that they do not present KAM behavior), they become finer in the
transition to chaos  when
using measures
$\mcN_{\mcD}^{\max}$ and
$\mcN_{\mcD}^{\langle \cdot \rangle}$. Such features remain stable with respect to
the cutoff time, indicating that they are not random
fluctuations, and become finer as the dimension increases.\par
\begin{acknowledgments} 
We acknowledge the support by CONACyT and DGAPA-IN-111015,  as well useful discussions with Heinz-Peter Breuer, Diego Wisniacki and Thomas Gorin.
\end{acknowledgments}
\appendix*
\section{Dynamical regimes} 
\label{sec:dynamicregimes}
The spin chain has well known dynamical regimes in the sense of random matrix
theory. The analysis of the spectra (the eigenphases of the Floquet operator)
has been done for the chaotic regime and for $16$ qubits
in Ref.~\cite{prosenycarlos}.

In this appendix we present a brief analysis for the
integrable regime for $12$ qubits and for completeness also for the chaotic and
mixed
regimes, following the aforementioned work.
In order to show the correspondence of the eigenphases of the Floquet
operator with the results of random matrix theory, we have to identify the
subspaces corresponding to the good quantum numbers of the system.
We then  compute the distribution of the distance among the nearest neighbor
eigenphases [named $P(s)$] in each symmetry sector. 
The homogeneous spin chain has a symmetry under translation of spins, \ie{} the
Hamiltonian remains invariant if we take the spin $i$ to $i+1$. Thus we will
use the eigenspectra corresponding to the eigenspaces of the translation
operator $T$ for the analysis of $P(s)$. 

\begin{figure}
\centering
\includegraphics[scale=1]{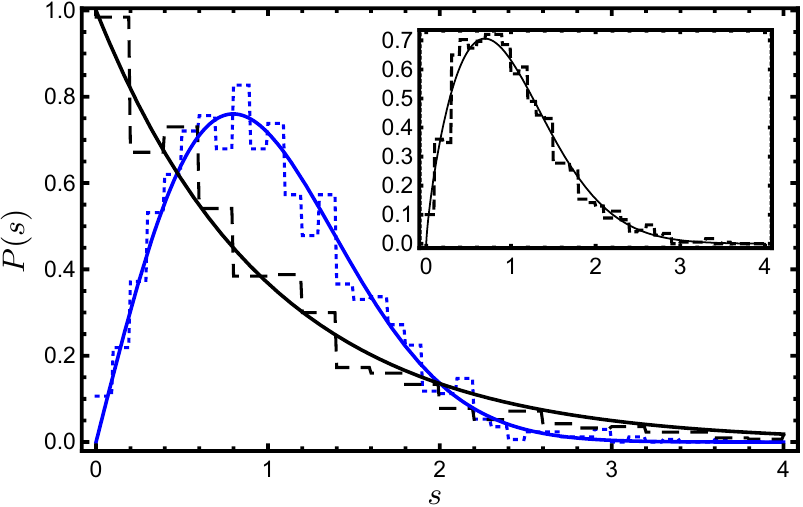}
\caption{The figure shows the nearest neighbor spacing distributions $P(s)$ of
the spin chain with $12$ qubits for two values of the control parameter. In the
main figure, the dotted blue curve shows the $P(s)$ for the chaotic regime, and
the
dashed black shows that for the integrable regime. The solid black curve shows the $P(s)$ for
the Poissonian orthogonal ensemble and the solid blue shows it for the circular
orthogonal ensemble. In the inset the dashed curve shows the $P(s)$ for the
mixed regime and the solid curve shows that for the Brody distribution with $b=0.77$; see
Ref.~\cite{Brody1981}. There is good agreement on all regimes 
with the predictions of random matrix theory.\label{fig:Ps}}
\end{figure}

The symmetry operator acts in the computational basis $\ket{\alpha_0,\dots,\alpha_{N-1}}$ ($\alpha_j \in \lbrace 0,1 \rbrace$), as $
T\ket{\alpha_0,\dots,\alpha_{N-1}}=\ket{\alpha_{N-1},\alpha_0,\dots,\alpha_{N-2}}$.
Since $T^N=\mathbb{I}$, its eigenvalues are simply $\exp\left( 2 \pi i k/N
\right)$ with $k$ an integer between 0 and $N-1$. Therefore, the Hilbert space
is foliated into $N$ subspaces $\mcH=\oplus_{k\in \mathbb{Z}_{/N}}\mcH_k$.
The chain also has a reflection symmetry given the symmetry operator $R$, which
transforms
$R\ket{\alpha_0,\dots,\alpha_{N-1}}=\ket{\alpha_{N-1},\dots,\alpha_{0}}$.
This symmetry commutes with the $T$ in the subspace identified by
$k=0$, and for even $N$, also in $k=N/2$; 
for simplicity these subspaces are removed from
the calculation. 
Figure~\ref{fig:Ps} shows the averaged nearest neighbor spacing distribution
over the relevant subspaces, and the ansatz corresponding to the different
dynamical regimes~\cite{Brody1981}.
For the integrable regime, we plot the Poisson distribution
$e^{-s}$; for the chaotic we plot
the Wigner surmise; finally, for the mixed regime,
we present the Brody distribution~\cite{Brody1973},
$$P_q(s)=(q+1) s^q \Gamma \left(\frac{q+2}{q+1}\right)^{q+1} e^{-s^{q+1} \Gamma \left(\frac{q+2}{q+1}\right)^{q+1}}.$$
The Brody parameter is denoted by $q$ and takes the ansatz from the integrable
case ($q=0$) to the Gaussian orthogonal ensemble ($q=1$), fitting smoothly with the nearest spacing
distribution of the chain in the transition to chaos.
\bibliographystyle{unsrt}
\bibliography{labibliografia}
\end{document}